# A Plainified Composite Absorber Enabled by Vertical Interphase

Yuhan Li, Faxiang Qin*, Le Quan, Huijie Wei, Yang Luo, Huan Wang, Hua-Xin Peng

Institute for Composites Science Innovation, Department of Materials Science and Engineering, Zhejiang University, Hangzhou, China

**ABSTRACT:** Interface constitutes a significant volume fraction in nanocomposites, and it requires the ability to tune and tailor interfaces to tap the full potential of nanocomposites. However, the development and optimization of nanocomposites is currently restricted by the limited exploration and utilization of interfaces at different length scales. In this research, we have designed and introduced a relatively large-scale vertical interphase into carbon nanocomposites, in which the dielectric response and dispersion features in microwave frequency range are successfully adjusted. A remarkable relaxation process has been observed in vertical-interphase nanocomposites, showing sensitivity to both filler loading and the discrepancy in polarization ability across the interphase. Together with our analyses on dielectric spectra and relaxation processes, it is suggested that the intrinsic effect of vertical interphase lies in its ability to constrain and localize heterogeneous charges under external fields. Following this logic, systematic research is presented in this article affording to realize tunable frequency-dependent dielectric functionality by means of vertical interphase engineering. Overall, this study provides a novel method to utilize interfacial effects rationally. The research approach demonstrated here has great potential in developing microwave dielectric nanocomposites and devices with targeted or unique performance such as tunable broadband absorbers.

# 1 INTRODUCTION

Interface has a crucial influence on the overall performance of composites both mechanically and functionally[1-3]. Especially in terms of dielectric properties, the permittivity, breakdown strength, dielectric dispersion and relaxation dynamics of composites are very sensitive to interfacial properties[4-7]. With the filler size decreasing to nanoscale, interfaces become more complex and spacious, playing a pivotal role in the overall dielectric functionality[8, 9]. Thereby, great attention has been paid to reveal the internal correlation between interface and dielectric properties both theoretically and experimentally[3, 10-12]. Maxwell-Wagner-Sillars effect gives a general explanation of interface-induced polarization, which originates from the differences in conductivity, permittivity and relaxation time of charge carriers in the materials across the interface[13-15]. By tuning interfacial properties with various chemical and engineering methods, the polarization ability and dielectric response of nanocomposites are also changed accordingly[16-20]. For example, the permittivity of carbon nanotube (CNT)/ polyvinylidene fluoride (PVDF) nanocomposites was reported to increase remarkably with enhanced molecular interaction and huge interfacial area[4]. In the microwave frequency range, dielectric relaxation peaks of CNT/silicone elastomer (SE) nanocomposites were shown to be sensitive to different types of interfacial interaction between chemically modified CNTs and polymer matrix[21]. While great attention has been given to explore these phenomena from various perspectives, limited research is carried out on utilizing the interfacial effects to tailor the dielectric properties rationally and effectively.

Based on the existed research that focuses on the filler-matrix interface, it would be promising if we can amplify these effects by enlarging the interfacial region to realize the true potential of interface engineering. Here, we propose "plainified composites" to indicate composites with optimized performance achieved by only interface engineering (Figure 1). As shown in the figure, canonical composites are normally composed of the matrix and the filler. Traditionally, certain properties of composites can be improved by adding more fillers or various types of fillers. However, the design/manufacture complexity and structural disintegrity will be increased accordingly. Meanwhile, the lightweight feature of composites could also be compromised when the filler loading becomes higher. In comparison, plainified composites are extremely potential as the material system and concentration of fillers remain unchanged. In these composites, interfacial effects can be fully utilized through various methods such as modifying filler-matrix interface, introducing in-built interphase or large-scale interface without the penalty of density and structural integrity. In metallic materials, the concept of "plainification" has also been highlighted, in which tailored microstructures are achieved by modifying or altering grain boundaries instead of changing the composition[22]. Thereby, the proposed research topic is of general interest towards a wide range of applications. Through engineering interfaces at different length scales, we will be able to realize more efficient design methodology and superior materials performance.

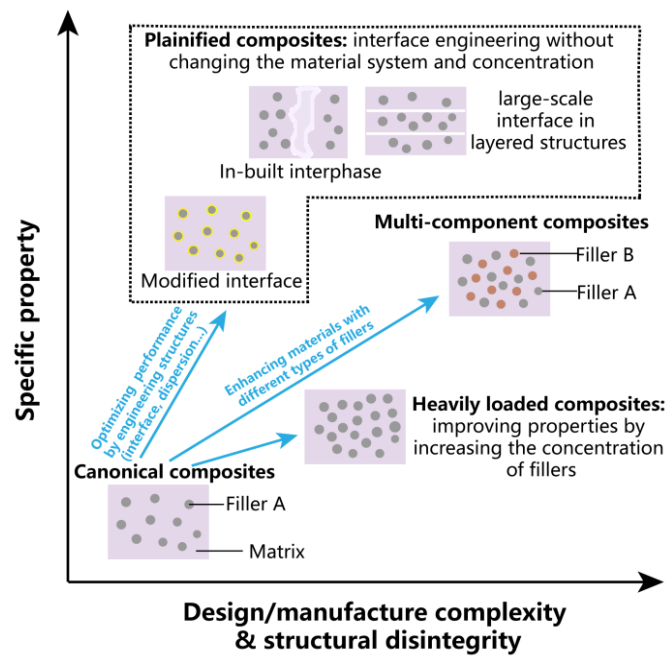

Figure 1. Plainified composites enabled by interface engineering (the $X$ axis represents the relative design/manufacture complexity and structural disintegrity, and the $Y$ axis represents the specific property of composites)

The implication of interface engineering and plainified composites can be magnificent in the context of dielectric functionality. Previous research has shown that the interfacial area of composites has a major influence on the dielectric performance. In storage capacitors, by building multi-layer or sandwich structures, the interfacial area can be greatly increased[5]. The broad interfacial area between layers could restrict the tunneling effect of electrons and delay the breakdown of materials under high electric field. Joyce et.al introduced multi-layer structure into polymer capacitors and increased the energy barrier at the interface, thereby minimizing tunneling current and increasing the energy density of the capacitor effectively[23]. Meanwhile, there are charge accumulations and polarizations at the interface between layers, enhancing the dielectric response and improving the permittivity in a great degree[24]. Such kind of effect is similar to the interfacial polarization that is originated from the distinct electrical properties between filler and the matrix. It will introduce new polarization mechanisms and relaxation processes into the whole system and change the overall dielectric response. Moreover, the large-scale interface in layered structures opens up new possibilities as it can be easily engineered through varying the composition, arrangement, and thickness of each layer. For instance, Wang et.al have designed bilayer high-$k$ composites by graphite/polyvinylidene fuluoride (PVDF) composites with positive and negative $\varepsilon$ respectively[25]. The permittivity of the bilayer samples can be easily tuned in a broad range by adjusting the filler contents and the thickness of the two layers. Likewise, in multi-layer structures composing of alternative PVDF and carbon black/PVDF layers, gradual enhancement of permittivity in the frequency range of $10^2$-$10^7$ Hz can be realized by increasing the number of layers[26]. The position and intensity of the multilayer structure induced loss peak between $10^3$-$10^4$ Hz also change upon layer multiplication. In this respect, the interface in these structures realize the accumulation and confinement of heterogeneous charges at a higher length scale, which can be further developed to tune the dielectric functionality stably and reliably.

To utilize interfacial effects rationally, it is also important to consider the relative position of the interface and the propagating direction of electromagnetic waves to generate effective interaction between them. In multi-layer dielectric functional materials, the relative direction between the layered structure and the electric field plays a decisive role on dielectric properties. According to Teirikangas et al, in the 'horizontal structure', the distribution of the electric field is relatively homogeneous along the interface, while in the 'vertical structure' this continuity is broken by the interface[27], resulting in distinct dielectric response. In multiferroic oxide heterostructures, vertical interface has also been explored to manipulate the electromagnetic properties. In this context, the hetero-interface is vertical to the substrate surface, reducing the influence of the substrate and increasing strain tenability at the same time. Hence, it is promising for precise control of mechanical and electromagnetic properties[28]. To this end, large-scale interfaces that have effective interaction with external fields will be critical for manipulating the overall performance of dielectric functional nanocomposites.

In this paper, we take inspiration from the above perspectives and introduce a vertical interphase in carbon nanotube/silicone elastomer nanocomposites to fully explore interfacial effects and expand the tunability of electromagnetic properties (Figure 2). The vertical interphase is composed of nanocomposites with different polarization abilities across the interphase. The interphase can function by accumulating charges, restraining their local distribution and inducing dielectric response that is distinguishable from that of homogeneous materials. The proposed method has great potential in enhancing the high frequency dielectric response of nanocomposites and enables the full utilization of microscopic and macroscopic interfacial properties. It provides new insights for the design and fabrication of lightweight nanocarbon microwave absorbing materials and other microwave functional materials. By introducing vertical interphase into nanocomposites, this research brings the investigation of interface related dielectric response to a higher level and takes full advantage of interfacial engineering in dielectric functionality context.

## 2 EXPERIMENTAL SECTION

### 2.1 Materials

Multi-walled carbon nanotubes (outer diameter: <8 nm, inner diameter: 2-5 nm, length: 10-30 μm) were obtained from Chengdu Organic Chemicals Co., Ltd, Chinese Academy of Sciences. The nanotubes were grown by Chemical Vapor Deposition method and had a purity of 95%. SYLGARD(R) 184 silicone elastomer kit (Dow Corning) was used as the polymer matrix of carbon nanocomposites. Sodium nitrate ($NaNO_3$), potassium permanganate ($KMnO_4$), and sulfuric acid ($H_2SO_4$) were purchased from Sinopharm Chemical Reagent Co., Ltd. Dopamine hydroxide was supplied by Aladdin Co., Ltd. γ-Methacryloxypropyl trimethoxy silane (KH570, coupling agent) was purchased from Adamas-beta.

### 2.2 Design of vertical-interphase nanocomposites

To introduce a large-scale interphase region that could effectively interact with electromagnetic waves, vertical-interphase carbon nanocomposites are designed as follows (Figure 1). The relative position of the sample and the incident microwave is shown in Figure 1a. The distribution of electric field under TE$_{10}$ mode is shown in Figure 1b. Figure 1c is the schematic description of CNT/SE nanocomposites. Various surface and interface modification methods are applied to achieve different polarization abilities in CNT/SE nanocomposites. A relatively broad interphase area is formed with two kinds of premixed CNT/SE nanocomposites through flow and diffusion (Figure 1d). As such, the relatively broad interphase region is introduced and expanded along the *Z* direction so that it can fully interact with microwave (distributed in the *XY* plane).

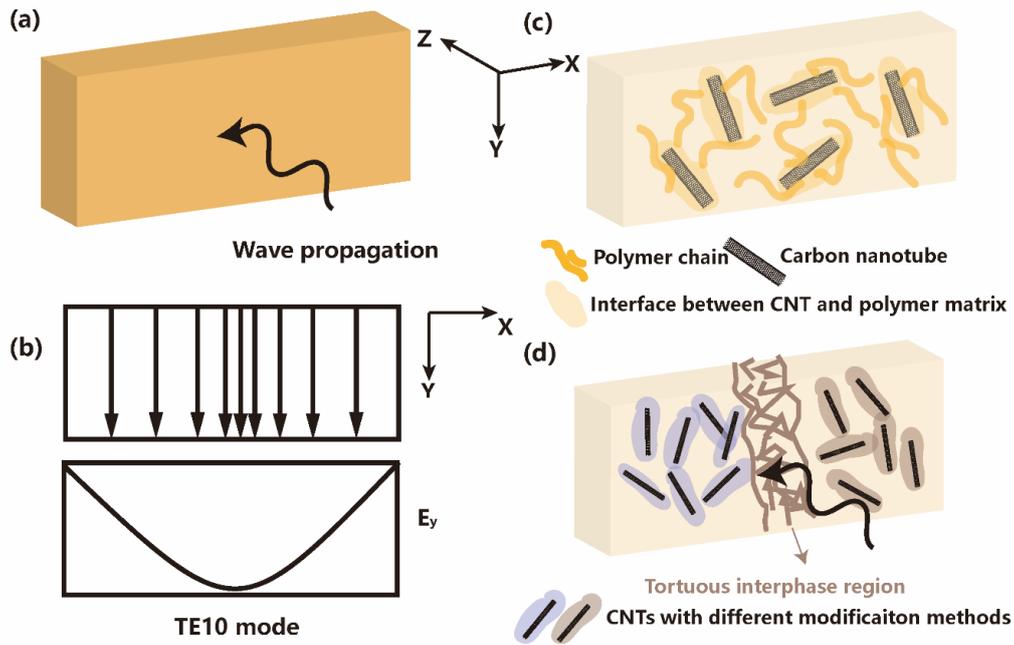

**Figure 2.** Schematic design of vertical-interphase nanocomposites: (a) sample for waveguide measurements and the corresponding incident direction of electromagnetic waves (b) direction and distribution of electric field under TE$_{10}$ mode (c) schematic description of CNT/polymer nanocomposites (d) building vertical interphase in CNT nanocomposites with different CNT modification methods

**2.3 Modification strategies towards nanocomposites with different polarization ability**

In order to prepare nanocomposites with different polarization abilities, several CNT surface modification methods were chosen. Silane coupling agent (KH570) modified CNT is marked as CACNT. As-received CNTs were first dispersed into KH570/ ethanol solution (1 wt%). The solution was sonicated for 1 h before drying in oven under 60 °C. Oxidization of CNT was carried out with strong oxidants (KMnO$_4$, H$_2$SO$_4$). 0.5 g CNT was mixed with 0.375 g NaNO$_3$, 1 g KMnO$_4$ and dispersed in 15 ml H$_2$SO$_4$ under room temperature and stirred for 24 h. 50 ml H$_2$O and 3 ml H$_2$O$_2$ were then added. The mixture was washed with deionized water and dried under 80 °C. The product of chemical oxidation is marked as OCNT. Further modification was achieved by mixing OCNT with dopamine hydrochloride aqueous solution (2 g/L) and stirring for 10 h under 60 °C. The final product is marked as DPACNT.

**2.4 Preparation of vertical-interphase nanocomposites**

The solution mixing method was utilized to prepare premixed CNT/SE nanocomposites. CNTs were dispersed in tetrahydrofuran (THF) and silicone elastomer was also dissolved in THF simultaneously, followed by mixing in a planetary centrifugal mixer. The solvent was then evaporated and the premixed nanocomposites were degassed for use (marked as CNT/SE, CACNT/SE, OCNT/SE, DPACNT/SE depending on the type of nanofillers). For each vertical-interphase nanocomposites, two kinds of premixed nanocomposites were prepared at the same time (e.g. CNT/SE and OCNT/SE). Equal amounts of CNT/SE and OCNT/SE were poured into the mold from two sides. The premixed nanocomposites flowed and diffused from both sides of the mold and blended in the central area, forming an interphase region. The vertical-interphase nanocomposites were then cured at 125 °C. The same method was used to prepare a set of nanocomposites with different compositions. The dimension of the mold is 22.86 mm×10.16 mm×2 mm.

**2.5 Characterization**

A field emission scanning electron microscope (Zeiss, Utral 55) was used for observing the morphologies of the samples. The dispersion and distribution of CNTs were monitored by an optical microscope (Olympus BX53M). *ImageJ* (an open-source software) is applied to present the skeletonized pictures of optical images and carry out the statistical analysis of the average area of CNT aggregates to better illustrate the structure of the interphase. The chemical structures of CNTs and modified CNTs were characterized by Fourier transform infrared spectroscopy (FTIR, ThermoFisher). Raman spectroscopy (DXR smart Raman spectrometer, irradiation wavelength: 532 nm) was performed for the nanocomposites. A vector network analyzer (R&S, ZNB20) was used to measure the scattering parameters. The complex permittivity in the frequency range of 8.2-12.4 GHz (*X* band) was extracted by Nicolson-Ross-Weir method. *1stopt* (an optimization software, developed by 7D-Soft High Technology Inc.) was used for curve fittings and the extraction of characteristic relaxation times.

**3 RESULTS AND DISCUSSION**

**3.1 Carbon nanocomposites with different polarization abilities**

Various chemical modification methods are first explored to achieve different polarization abilities for building vertical-interphase in carbon nanocomposites. Silane functionalization of CNT could improve its interaction with the silicone elastomer. The siloxy group on KH570 molecules would interact with the oxygen-containing groups on CNTs through hydrolysis and condensation, and the vinyl group can participate in the vulcanization reaction of silicone rubber[29]. Oxidation is commonly adopted to introduce oxygen functional groups on the surface and ends of CNT, which is helpful for improving its compatibility with polymer and reducing agglomeration[30, 31]. These functional groups such as carboxyl can further take part in a variety of chemical reactions and realize secondary modification and functionalization[32]. Dopamine has been considered to be extremely adhesive to various surfaces since 2007, when Lee et.al first used it for coating a wide range of materials[33]. Meanwhile, the characteristic self-oxidative polymerization of dopamine makes it especially suitable for surface modification. For instance, dopamine modification was shown to improve the dispersion of $TiO_2$ nanofibres in PVDF and mitigate the concentration of electric field efficiently by forming a shell around the fillers[34].

The chemical structures of unmodified and modified CNTs were investigated by FTIR spectroscopy as shown in Figure 3a. Comparing to raw CNTs, there appear absorption peaks at 933 $cm^{-1}$, 1011 $cm^{-1}$, and 1300 $cm^{-1}$ for CACNT that can be attributed to C=C stretch vibrations, Si-O stretch vibrations, and C-O-C stretch vibrations from KH570 molecules[29]. The oxidation process introduced the characteristic peak at 1712 $cm^{-1}$, corresponding to stretch vibrations of C=O from carboxyl groups. Meanwhile, there is a significant peak at 2350 $cm^{-1}$ that can be assigned to vibrations of hydrogen bonding, suggesting the strong interaction between the increased oxygen-bearing groups. Further modification of dopamine is evidenced by the appearance of peaks at 1427 $cm^{-1}$, 1576 $cm^{-1}$ attributed to aromatic C-C and N-H stretch vibrations respectively[34]. The modified nanotubes were then incorporated into silicone elastomer to fabricate nanocomposites. The Raman spectra of the as-prepared carbon nanocomposites are presented in Figure 3b. The D bands and G bands are observed for all samples at 1342 cm$^{-1}$ and 1592 cm$^{-1}$ respectively, which represent the disorder-induced double-resonance and in-plane vibrations of sp$^2$ C-C bonds[31, 35, 36]. The relative intensity ratios of D peak to G peak ($I_D/I_G$) is used for estimating the degree of defects and the destruction of graphitic integrity. According to Figure 3b, the increases of $I_D/I_G$ are all observed for OCNT/SE, CACNT/SE, and DPACNT/SE comparing to CNT/SE. Thereby, different chemical modification methods have all influenced the vibrations of carbon atoms and increased the degree of defects on CNT, which would result in nanocomposites with distinct microstructures and interfacial features.

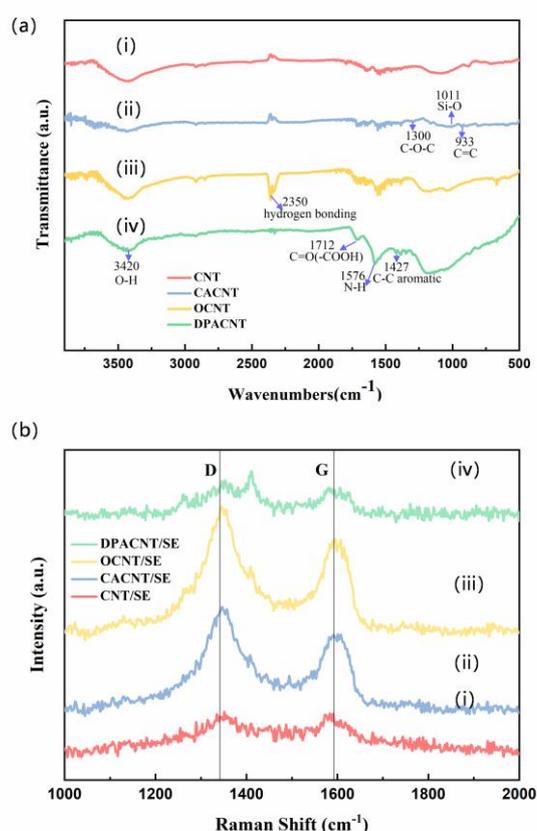

**Figure 3**. (a) FTIR spectra of CNTs modified by different methods (the appearance of characteristic peaks are indicated by blue arrows): (i) raw CNTs, (ii) CNTs modified with KH570, (iii) oxidized CNTs, (iv) CNTs further functionalized with dopamine; (b) Raman spectra of CNT/silicone elastomer nanocomposites with different modification methods (the positions of D peak and G peak are indicated by the vertical lines): (i) CNT/SE, (ii) CACNT/SE, (iii) OCNT/SE, (iv) DPACNT/SE

The SEM images of CNT/silicone elastomer nanocomposites with different modification methods are shown in Figure 4. From Figure 4a, e, i, it is observed that there exist many large-size agglomerates (about 10-20 μm) in CNT/SE. In comparison, silane functionalization has improved the dispersion of nanotubes, decreasing the agglomerates size to below 10 μm (Figure 4b, f, j). In both OCNT/SE and DPCNT/SE nanocomposites, the dispersions of CNTs are greatly enhanced. The agglomerates appear to be dotted-like in Figure 4c-d, and the distributions of CNTs are very uniform. Under higher magnification, it could be seen that DPACNTs are more homogeneously dispersed in the matrix than OCNTs (Figure 4k, i). Thereby, the relative dispersion abilities of the nanotubes in silicone elastomer are as follows: DPACNT>OCNT>CACNT>CNT. Silane functionalization is not as effective as oxidation in reducing agglomeration because there are limited active sites on raw CNT. Meanwhile, wrapping OCNT with polydopamine has further improved the dispersion degree. Thus, four distinct dispersion states are achieved in CNT nanocomposites by diverse chemical modification methods and different degrees of modification.

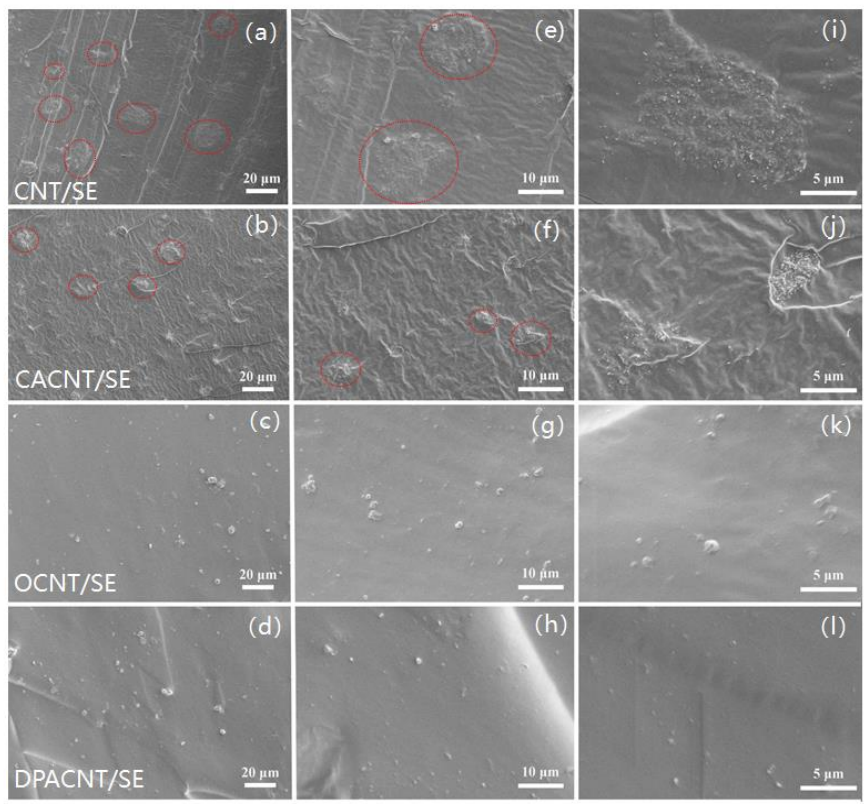

**Figure 4**. SEM images of CNT/silicone elastomer nanocomposites ($f$=0.5 vol%): (a, e, h) CNT/SE; (b, f, i) CACNT/SE; (c, g, h) OCNT/SE; (d, h, i) DPACNT/SE (CNT agglomerates are indicated by red circles)

The differences in the chemical structures of nanofillers and the microstructures of nanocomposites have resulted in distinct dielectric properties, which are illustrated as follows. In Figure 5, the imaginary permittivities of the above CNT nanocomposites are plotted as a function of filler loadings under selected frequencies. It is observed in the first place that the imaginary permittivities for all the nanocomposites are all very small and show little dependence on frequency. For each loading, various modification methods have resulted in different degrees of decrease in permittivity. Generally, the $\varepsilon''$ for each filler loading follows the trend of $\varepsilon''_{CNT/SE} > \varepsilon''_{CACNT/SE} > \varepsilon''_{OCNT/SE} > \varepsilon''_{DPACNT/SE}$ as a result of different modification and functionalization mechanisms. These chemical treatments influenced not only the conductivity of CNTs but also their dispersion ability and interfacial interaction with polymer matrix.

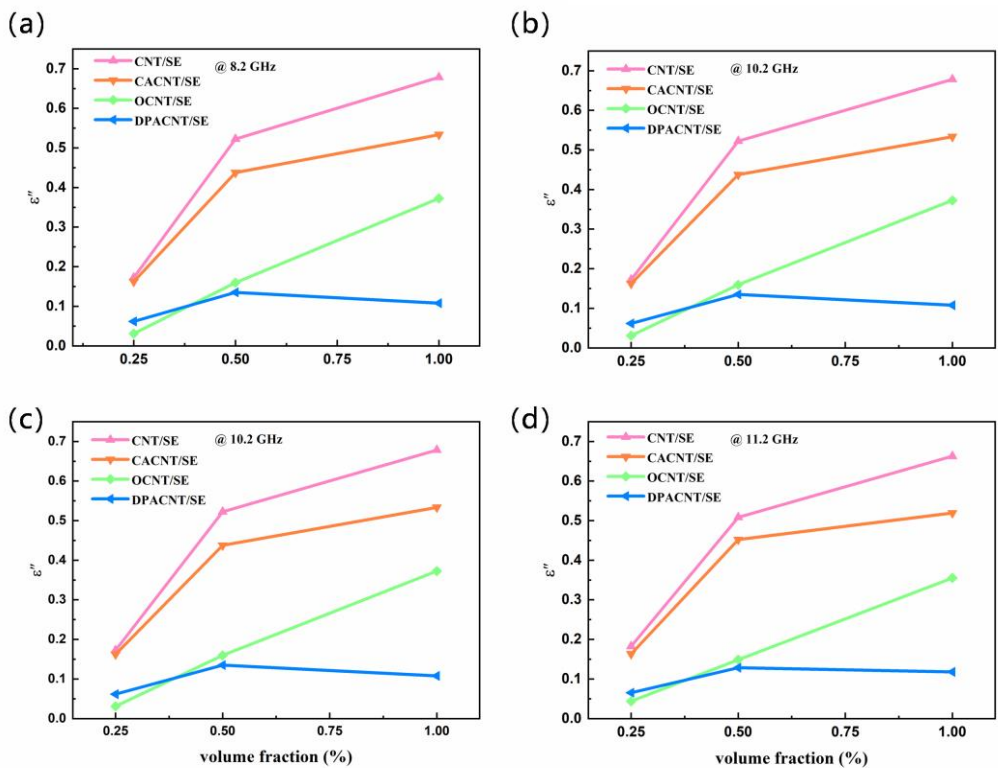

**Figure 5.** Imaginary permittivities of CNT/silicone elastomer nanocomposites with different modification methods and filler loadings: (a) 8.2 GHz; (b) 9.2 GHz; (c) 10.2 GHz; (d)10.2 GHz

To be specific, while silane functionalization partially affected the original structure of CNTs through chemical bonding, the decrease of permittivity is relatively moderate due to limited active sites on raw CNTs. When CNTs were treated with strong oxidants, the structure of nanotubes was severely destroyed and the conductivity was greatly compromised. Thereby, the $\varepsilon''$ of OCNT/SE and DPACNT/SE decrease to below 0.1 ($f$=0.25 vol%). It should be noticed that when OCNTs were wrapped with polydopamine, the $\varepsilon''$ of nanocomposites further drops because polydopamine functions as a protective shell restricting the dielectric loss at the interface. At this point, the permittivity does not change with filler loading anymore, indicating that the dielectric response becomes very insignificant. Except for the original structure of CNTs, the dispersion ability also plays a role in influencing the dielectric properties. When the functional fillers are better dispersed, there are less local conductive networks and the overall dielectric response can also be decreased. As such, the conductivity and the dispersion of nanofillers are the primary causes of the ultimate differences in dielectric properties. Real permittivities of the samples as function of filler loading are presented in Figure S1 (Supporting Information), in which a similar pattern is observed, again confirming the effectiveness of these methods in adjusting dielectric properties. To this end, a set of nanocomposites with distinct polarization abilities are designed and fabricated.

**3.2 Vertical interphase induced dielectric relaxation**

CNT/SE nanocomposites developed in 3.1 are used for building vertical-interphase nanocomposites. The vertical-interphase nanocomposites are fabricated by diffusion and partial mixing of premixed nanocomposites and are named correspondingly. For example, the vertical-interphase nanocomposite constituted by CNT/SE and OCNT/SE is marked as CNT-OCNT/SE. The original optical microscope images of CNT-OCNT/SE are shown in Figure S2 (Supporting Information). The left side of the sample (OCNT/SE rich) is obviously much better dispersed than the right side (CNT/SE rich) due to the difference in dispersion states. The mixed region could be observed from the skeletonized image of the central area (Figure 5a). Statistical analyses on the average area of agglomerates are carried out with *ImageJ* for the left (L), central (C), and right (R) area of the sample respectively as a relative indication of dispersion degree[37, 38]. Each region is divided into 6 sections and the results are presented in Figure 6b, in which L and R represents the overall estimation of the left region and the right region. In comparison, the average area of agglomerates in C1, C2, and C3 are similar to that of the L region. The dispersion is much worse in C4, C5, and C6, indicating that these sections are mainly constituted by CNT/SE. There is sharp increase in the average area of CNT agglomerates from C3 to C4, suggesting that the interphase is mainly introduced around these two sections. While there is no clear boundary, the interphase is supposed to be a relatively large area with a torturous path forming along the *Y* and *Z* direction.

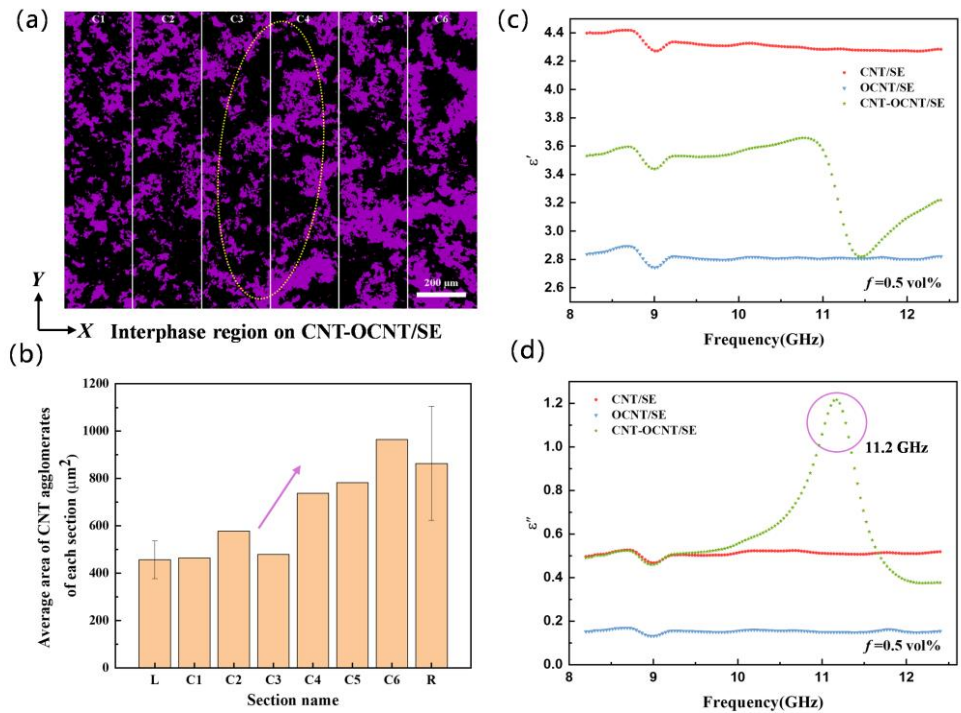

**Figure 6**. Optical microscope images analyses of CNT-OCNT/SE nanocomposites (*f*=0.5 vol%): (a) skeletonized optical image of the interphase region (marked by the dashed oval); (b) statistical analysis of the average area of CNT aggregates on each section (the purple arrow indicates the abrupt change of agglomerates area); (c, d): frequency dependence of complex permittivity of CNT nanocomposites with or without in-built vertical interphase: (c) real part; (d)imaginary part;

Frequency dependence of permittivity is used for analyzing the dielectric properties of vertical-interphase nanocomposites. For CNT-OCNT/SE nanocomposites, the complex dielectric spectra are exhibited in Figure 6c-d. The $\varepsilon'$ and $\varepsilon''$ of CNT/SE and OCNT/SE are relatively stable over the whole *X* band and there is a distinct discrepancy in their polarization abilities. The $\varepsilon'$ of CNT-OCNT/SE is in between that of CNT/SE and OCNT/SE (Figure 6c). It first remains steady as the frequency increases, and then starts to drop continuously over 10.8-11.5 GHz. Simultaneously, a remarkable relaxation peak appears in the imaginary spectra of CNT-OCNT/SE over this frequency range (Figure 6d). The peak value of $\varepsilon''_{CNT\text{-}OCNT/SE}$ reaches 1.2 at 11.2 GHz, which is significant comparing to $\varepsilon''_{CNT/SE}$ (0.5) and $\varepsilon''_{OCNT/SE}$ (0.15) at the same frequency. Such variation trend in complex dielectric spectra of CNT-OCNT/SE represents a typical dielectric relaxation process. It can be logically concluded that this process is closely correlated with the introduction of vertical interphase, which induces strong and effective interaction between the nanocomposite and electromagnetic waves.

**3.3 Tunable microwave dielectric response enabled by vertical interphase**

To further reveal the effect of introducing vertical interphase into nanocomposites, we built several vertical-interphase nanocomposite systems with varied compositions and studied their dielectric response under external fields. These systems are labeled as 1, 2, 3, 4 and 5 according to the discrepancy in polarization ability of the two regions across vertical interphase ($\Delta\varepsilon$). It should be noted that the label of each system is only a comparative description marking the difference in dielectric properties of the premixed nanocomposites. Considering that $\Delta\varepsilon$ should be frequency-dependent, we could not specify it as an absolute value. Figure 7a presents the imaginary dielectric spectra of nanocomposites with a single type of modified or raw CNTs. These nanocomposites are overall homogeneous and their imaginary permittivities are relatively stable over $X$ band. We first fabricated vertical-interphase nanocomposites that are composed by OCNT/SE and DPACNT/SE, which have the smallest discrepancy in polarization ability. It can be seen in Figure 7b that the dielectric response of OCNT-DPACNT/SE is very weak ($\varepsilon''<0.2$) and there is no obvious change in its dielectric spectra comparing to homogeneous nanocomposites. As $\Delta\varepsilon$ increases, there appears a weak relaxation peak in the dielectric spectra of CNT-CACNT/SE (at 10.7 GHz) as shown in Figure 7c. For OCNT-CACNT/SE, a stronger relaxation peak is observed at 11.3 GHz, accompanied by the peak value of $\varepsilon''$ increasing to around 0.8. Meanwhile, the blue shift of relaxation peak can be attributed to improved dispersion state and reduced relaxation time with the addition of OCNT. In this context, the $\Delta\varepsilon$ of vertical-interphase nanocomposites is the key to the appearance and enhancement of the characteristic relaxation peak. We then enlarged $\Delta\varepsilon$ and built system 4 and 5, in which significant changes of dielectric spectra are observed and the peak values of $\varepsilon''$ increase remarkably to 1.2 and 1.1 (Figure 7d). Simultaneously, these relaxation peaks at around 11.2 GHz become extremely sharp. It is thereby indicated that the relaxation process is enhanced greatly with increasing $\Delta\varepsilon$, again validating that it is originated from the difference in dielectric properties between the two regions across interphase. It is also noteworthy that the position and the intensity of the relaxation peak remain almost unchanged due to limited increase in Δ$\varepsilon$ from CNT-OCNT/SE to CNT-DPACNT/SE.

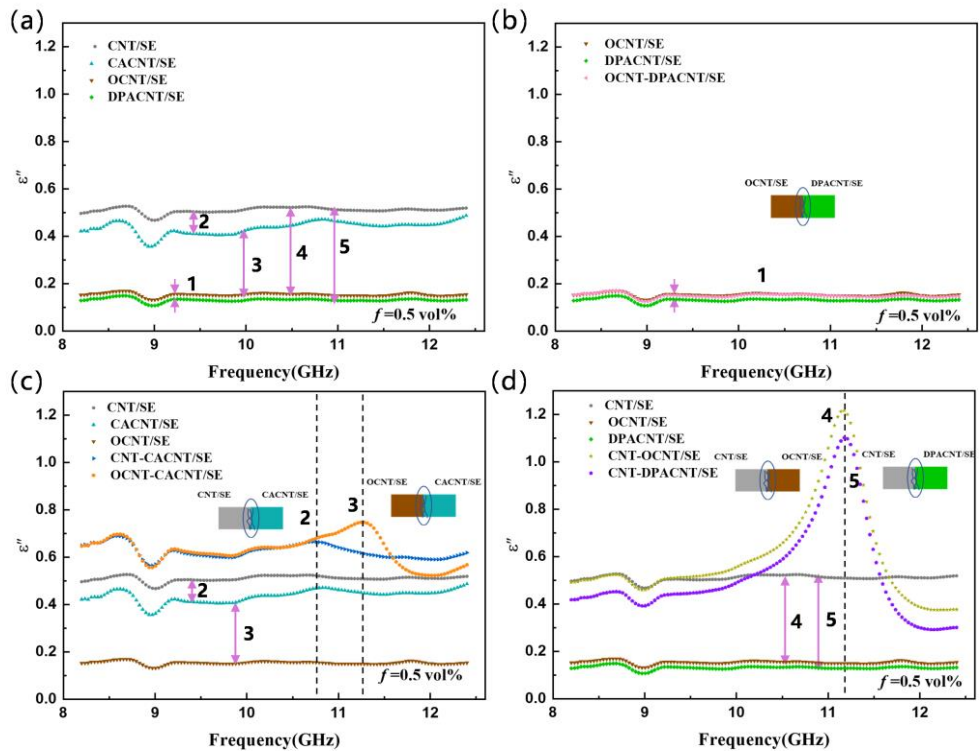

**Figure 7**. Frequency dependence of imaginary permittivity of CNT/silicone elastomer nanocomposites ($f$=0.5 vol%) for studying vertical interphase (8.2-12.4 GHz): (a) carbon nanocomposites with a single type of modified or raw CNTs (Δ$\varepsilon$ marked by the purple double headed arrows); (b) OCNT-DPACNT/SE vertical interphase nanocomposites; (c) CNT-CACNT/SE and OCNT-CACNT/SE nanocomposites; (d) CNT-OCNT/SE and CNT-DPACNT/SE nanocomposites (the insets: the schematic descriptions of different vertical-interphase nanocomposite systems, and various premixed nanocomposites are marked by different colors)

We further plot the frequency dependence of imaginary permittivity of vertical-interphase nanocomposites ($f$=0.5 vol%) in Figure 8a to investigate the intrinsic mechanism of dielectric enhancement and the new relaxation process enabled by vertical interphase. It is clear that the relaxation peak varies significantly with increasing Δ$\varepsilon$ (from 1 to 5). The Cole-Cole plots of these samples are exhibited in Figure 8b correspondingly. Firstly, the permittivity of OCNT-DPACNT/SE is very small and its dispersion feature is not very obvious. Secondly, the shape of the Cole-Cole plots for CNT-CACNT/SE and OCNT-CACNT/SE change to be partially arc-like, reflecting the appearance of a typical relaxation process. Finally, when the $\Delta\varepsilon$ is raised to 4 or 5, the radius of the arc increases astoundingly to almost circular, indicating that the $\varepsilon''$ experiences fierce changes over the investigated frequency range. Based on the above analyses, this characteristic relaxation process is originated from the polarization induced by the interphase, as shown in the inset of Figure 8b. To be specific, as there exist differences in the conductivity and permittivity between the left and right region of the sample, the vertical interphase becomes the centre of charge accumulation and restriction. A new polarization mechanism is induced and the relaxation occurs under alternating electric fields. While each vertical-interphase system has a different $\Delta\varepsilon$, its ability to restrict and accumulate charges are varied, enabling tunable dielectric response through controlling the compositions.

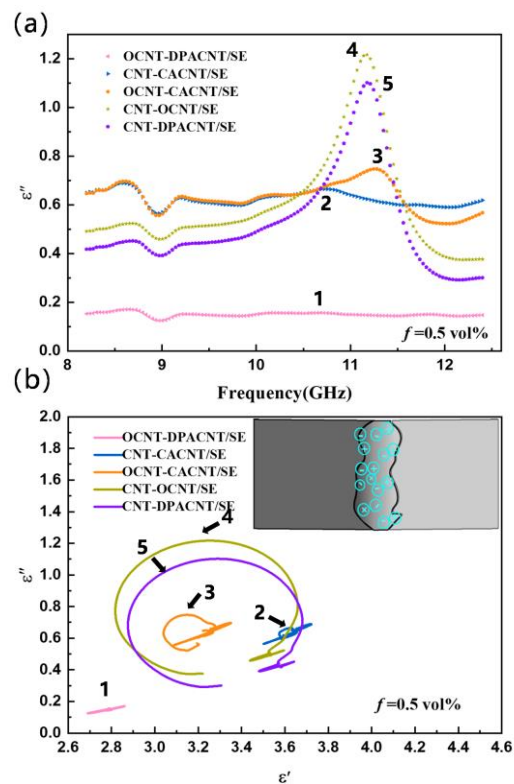

**Figure 8**. (a) Frequency dependence of imaginary permittivity of CNT/silicone elastomer nanocomposites ($f$=0.5 vol%) with in-built vertical interphase (8.2-12.4 GHz); (b) Cole-Cole plots and schematic description of the polarization mechanism (the inset) of the investigated nanocomposites

## 3.4 Comprehensive assessment of the vertical-interphase nanocomposites

According to the analyses above, the vertical interphase-induced relaxation relies on the discrepancy in polarization ability inside the nanocomposites. The filler loading of nanocomposites has an intrinsic influence on dielectric properties, so we focus on revealing the effect of vertical interphase in CNT nanocomposites as a function of different filler loading in this section. A relatively low filler loading (0.25 vol%) and a high one (1 vol%) are chosen to present a comprehensive assessment of the dielectric functionality in vertical-interphase nanocomposites. The frequency dependence of imaginary permittivity of CNT nanocomposites ($f$=0.25 vol%) is presented in Figure 9a. As the overall dielectric response is very weak ($\varepsilon''$< 0.3), we only display the dielectric spectra of CNT-OCNT/SE and CNT-DPACNT/SE here. They are marked as system 4* and 5* correspondingly. The $\varepsilon''$ of vertical-interphase nanocomposites are very similar to that of homogeneous nanocomposites, and weak relaxation peaks appear at 11.7 GHz.

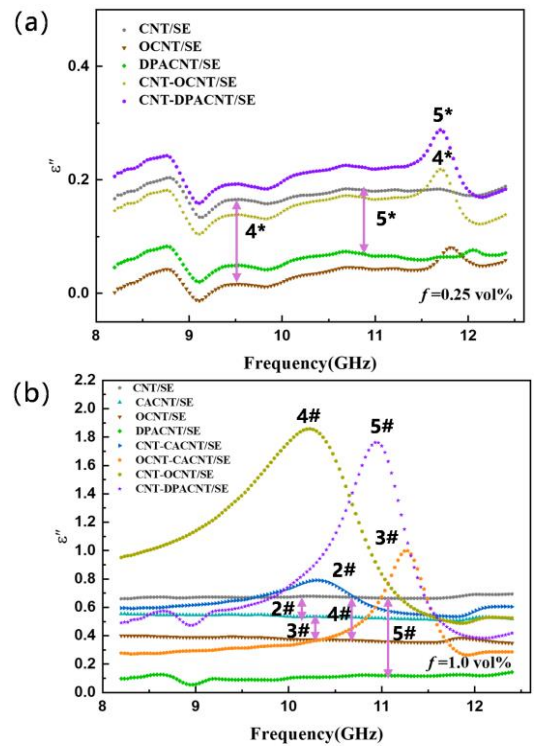

**Figure 9**. Frequency dependence of imaginary permittivity of CNT/silicone elastomer nanocomposites (with different filler contents) for studying vertical interphase ($\Delta\varepsilon$ marked by the purple double headed arrows): (a) $f$=0.25 vol%; (b) $f$=1 vol%

The imaginary dielectric spectra of CNT nanocomposites ($f$=1 vol%) are shown in Figure 9b. Four systems are built and labeled as 2#, 3#, 4#, and 5#. It can be observed that there appears a significant dielectric relaxation peak (between 9.5-11.5 GHz) for each vertical-interphase sample under this filler loading, suggesting that the introduction of vertical interphase induces a new relaxation process. Unlike homogeneous nanocomposites, the $\varepsilon$" of these samples vary remarkably with frequency. It is noteworthy that the position and intensity of the relaxation peak are distinct for different systems. The relaxation peak shifts to higher frequency when the overall dispersion of CNT is improved, contributing to a shorter relaxation time. Since the dispersion degrees of the nanocomposites are as follows: DPACNT/SE> OCNT/SE> CACNT/SE> CNT/SE, blue shifts of the relaxation peak are observed in 3# (to 11.2 GHz) and 5# (to 11.3 GHz) comparing to 2# and 4# respectively. To better illustrate this relaxation process, the real dielectric spectra are fitted by Cole-Cole equation[24, 39], which takes the form of:

$$\varepsilon = \varepsilon_\infty + \frac{\varepsilon_s - \varepsilon_\infty}{1+(i\omega\tau_0)^{1-\alpha}} \qquad (1)$$

in which $\varepsilon_\infty$ is the permittivity at high frequency limit, $\varepsilon_s$ is the static permittivity, $\tau_0$ represents the characteristic relaxation time and $\alpha$ represents the dispersion of relaxation time. The fitted curves are shown in Figure S3 (Supporting Information). The characteristic relaxation times of 2#, 3#, 4# and 5# are extracted and displayed in Table 1. The characteristic relaxation times of 3# and 5# are smaller than that of 2# and 4#, corresponding to the above analyses very well. Interestingly, although better dispersion is expected in 4# comparing to 2#, the relaxation time is not decreased, which is the result of stronger relaxation achieved by the enhanced ability of the interphase to restrict charges. In other words, the relaxation times are not only dependent on the overall dispersion of the sample, but are related to the interphase-induced relaxation process itself. In addition, the relaxation peak becomes more significant with increasing $\Delta\varepsilon$ (from 2# to 3#), accompanied by the peak value increasing from 0.8 to 1.0. CNT-OCNT/SE shows the strongest relaxation peak with the $\varepsilon$" reaching 1.8 at 10.2 GHz. When the $\Delta\varepsilon$ is further increased from 4# to 5#, the peak value of $\varepsilon$" remains almost stable and only the blue shift of the relaxation peak is observed. In this case, the charge accumulation at the vertical interphase reaches its limit at 4#.

**Table 1.** Fitted characteristic relaxation time of vertical-interphase nanocomposites

($f$=1 vol%)

| Sample | $\tau_0$ |
|---|---|
| 2#: CNT-CACNT/SE | 1.57e-11 |
| 3#: OCNT-CACNT/SE | 1.42e-11 |
| 4#: CNT-OCNT/SE | 1.58e-11 |
| 5#: CNT-DPACNT/SE | 1.46e-11 |

Comparative analyses are carried out to further display the ability of vertical interphase in tuning microwave dielectric response. The imaginary dielectric spectra of CNT-OCNT/SE and CNT-DPACNT/SE nanocomposites are plotted as a function of filler volume fraction in Figure 10a. With increasing filler loading, the nanofillers are worse dispersed, leading to the red shift of the characteristic relaxation peak. Simultaneously, the dielectric response is overall enhanced, which means that for the same composition, there is a larger discrepancy in the polarization ability across the interphase. As a result, the dielectric relaxation becomes even more predominant in nanocomposites with a higher filler loading. This phenomenon is proved by the Cole-Cole plots depicted in Figure 10b, in which circular curves with larger radius are observed upon increasing filler volume fraction. In this sense, it provides a new dimension to adjust the dielectric response within vertical-interphase nanocomposites.

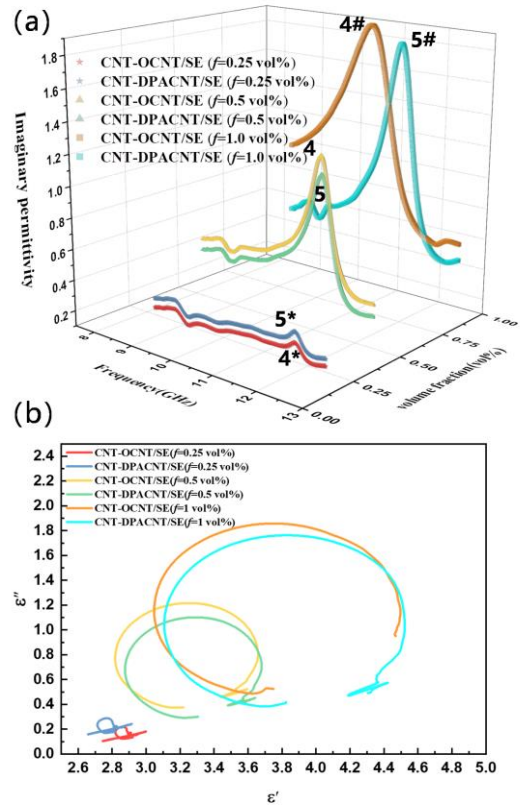

**Figure 10**. (a) Frequency dependence of imaginary permittivity of CNT-OCNT/SE and CNT-DPACNT/SE nanocomposites with different filler contents; (b) Cole-Cole plots of CNT-OCNT/SE and CNT-DPACNT/SE nanocomposites

The idea of engineering vertical-interphase carbon nanocomposites falls rightly into the scope of plainified composites, as in this research, only interface/interphase engineering methods are adopted to tailor dielectric functionalities. The vertical interphase enabled plainified nano*f*composites can be used as single-layer microwave absorbers or dielectric layers regulating the functionality of multilayer absorbers. They could be widely applied as composing layers of novel multilayer microwave absorbers to adjust the working frequency. For instance, vertical-interphase nanocomposites with differences in relaxation peak can be assembled to achieve broad-band microwave absorption (Figure 11) without introducing additional functional fillers and compromising the lightweight characteristic of the structure. Also, such a multilayer structure can maintain the structural integrity as all layers are essentially of the same CNT nanocomposite nature. Thus, it would be highly efficient and of significant engineering implication.

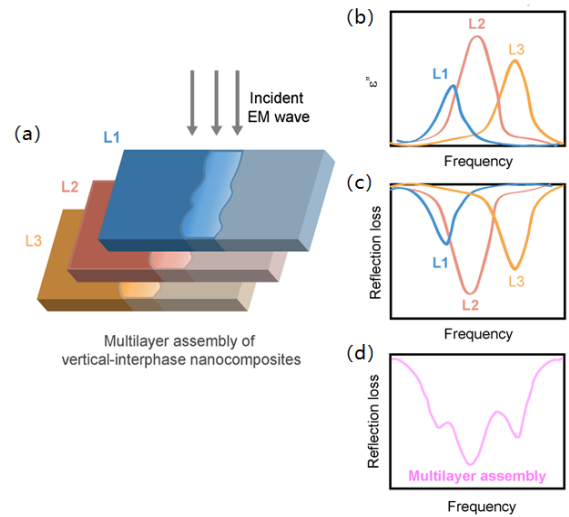

**Figure 11**. Multilayer assembly of vertical-interphase nanocomposites towards efficient microwave absorption: (a) Configuration of multilayer microwave absorber composed by vertical-interphase nanocomposites (L1, L2, and L3 represent layers with differences in the position of characteristic relaxation peak); (b) schematic representation of imaginary dielectric spectra for each layer; (c-d) schematic description of frequency dependence of reflection loss for each layer (c) and the multilayer assembly (d)

**4 CONCLUSIONS**

To summarize, this study presents a promising strategy to utilize interfacial effects to manipulate the dielectric response of nanocomposites facilely and effectively. On top of the typical filler-matrix interface, this study proposes the essential concept of plainified composites, in which only interface engineering methods are utilized to improve material properties. In the context of tuning dielectric functionality, a large-scale vertical interphase is rationally designed and successfully introduced into carbon nanocomposites in this study. The in-built vertical interphase becomes the centre of charge accumulation and confinement, triggering a strong and characteristic relaxation process in *X* band. The interphase-induced relaxation changes the dielectric dispersion pattern of carbon nanocomposites remarkably. Meanwhile, the intensity and position of the relaxation peak are highly tunable by adjusting the filler loading and the discrepancy in polarization ability across the interphase ($\Delta\varepsilon$). With large volume fraction of functional fillers or increasing $\Delta\varepsilon$, the relaxation process is greatly enhanced. From this perspective, it is convenient and efficient to tune the microwave dielectric properties of carbon nanocomposites. It is anticipated that stronger dielectric response will be achieved with higher loading of nanofillers, larger $\Delta\varepsilon$ enabled by other systems, or multiple interphase introduced in nanocomposites with a considerable $\varepsilon$ in future work. While the typical filler-matrix interfacial effect is sometimes insignificant due to small filler volume fraction, the ability to tailor dielectric response through vertical interphase without increasing the weight of materials is of great significance for microwave functional materials[40]. The insights provided here can be applied to reconfigurable microwave absorption to easily manipulate the microwave absorbing frequency. Overall, this study reveals an intrinsic dependence of dielectric functionality on interfacial properties at a higher length scale, thereby opening up new possibilities for the designing and engineering of microwave functional polymer nanocomposites and devices.

**Corresponding Author**

*Email: faxiangqin@zju.edu.cn

REFERENCES
(1) Coleman, J. N.; Khan, U.; Blau, W. J.; Gun'ko, Y. K., Small but strong: A review of the mechanical properties of carbon nanotube–polymer composites. *Carbon* **2006,** *44* (9), 1624-1652.
(2) Nan, C.-W.; Liu, G.; Lin, Y.; Li, M., Interface effect on thermal conductivity of carbon nanotube composites. *Appl. Phys. Lett.* **2004,** *85* (16), 3549-3551.
(3) Roy, M.; Nelson, J.; MacCrone, R.; Schadler, L. S.; Reed, C.; Keefe, R., Polymer nanocomposite dielectrics-the role of the interface. *IEEE Trans. Dielectr. Electr. Insul.* **2005,** *12* (4), 629-643.
(4) Yuan, J.-K.; Yao, S.-H.; Dang, Z.-M.; Sylvestre, A.; Genestoux, M.; Bai, J., Giant Dielectric Permittivity Nanocomposites: Realizing True Potential of Pristine Carbon Nanotubes in Polyvinylidene Fluoride Matrix through an Enhanced Interfacial Interaction. *The Journal of Physical Chemistry C* **2011,** *115* (13), 5515-5521.
(5) Wang, Y.; Cui, J.; Yuan, Q.; Niu, Y.; Bai, Y.; Wang, H., Significantly Enhanced Breakdown Strength and Energy Density in Sandwich-Structured Barium Titanate/Poly(vinylidene fluoride) Nanocomposites. *Adv Mater* **2015,** *27* (42), 6658-63.
(6) Sanida, A.; Stavropoulos, S. G.; Speliotis, T.; Psarras, G. C., Development, characterization, energy storage and interface dielectric properties in SrFe 12 O 19 / epoxy nanocomposites. *Polymer* **2017,** *120*, 73-81.
(7) Holt, A. P.; Bocharova, V.; Cheng, S.; Kisliuk, A. M.; White, B. T.; Saito, T.; Uhrig, D.; Mahalik, J. P.; Kumar, R.; Imel, A. E.; Etampawala, T.; Martin, H.; Sikes, N.; Sumpter, B. G.; Dadmun, M. D.; Sokolov, A. P., Controlling Interfacial Dynamics: Covalent Bonding versus Physical Adsorption in Polymer Nanocomposites. *ACS Nano* **2016,** *10* (7), 6843-6852.
(8) Lewis, T. J., Interfaces are the dominant feature of dielectrics at the nanometric level. *IEEE Trans. Dielectr. Electr. Insul.* **2004,** *11* (5), 739-753.
(9) Nelson, J. K., *Dielectric Polymer Nanocomposites*. Springer: New York, 2010.
(10) Wang, Y.; Zhang, Y.; Zhao, H.; Li, X.; Huang, Y.; Schadler, L. S.; Chen, W.; Brinson, L. C., Identifying interphase properties in polymer nanocomposites using adaptive optimization. *Compos. Sci. Technol.* **2018,** *162*, 146-155.
(11) Dang, Z.-M.; Yuan, J.-K.; Zha, J.-W.; Hu, P.-H.; Wang, D.-R.; Cheng, Z.-Y., High-permittivity polymer nanocomposites: Influence of interface on dielectric properties. *Journal of*

*Advanced Dielectrics* **2013,** *3* (03), 1330004.
(12) Sousa, A. A.; Pereira, T. A. S.; Chaves, A.; de Sousa, J. S.; Farias, G. A., Interfacial confinement in core-shell nanowires due to high dielectric mismatch. *Appl. Phys. Lett.* **2012,** *100* (21), 211601.
(13) Maxwell, J., Electricity and Magnetism, vol. 1, Clarendon. Oxford: 1892.
(14) Wagner, K., The after effect in dielectrics. *Arch. Electrotech* **1914,** *2* (378), e394.
(15) Sillars, R., J. Inst. Electr. Eng. **1937**.
(16) Kang, D.; Wang, G.; Huang, Y.; Jiang, P.; Huang, X., Decorating TiO2 Nanowires with BaTiO3 Nanoparticles: A New Approach Leading to Substantially Enhanced Energy Storage Capability of High-k Polymer Nanocomposites. *ACS Appl Mater Interfaces* **2018,** *10* (4), 4077-4085.
(17) Feng, Y.; Deng, Q.; Peng, C.; Hu, J.; Li, Y.; Wu, Q.; Xu, Z., An ultrahigh discharged energy density achieved in an inhomogeneous PVDF dielectric composite filled with 2D MXene nanosheets via interface engineering. *Journal of Materials Chemistry C* **2018,** *6* (48), 13283-13292.
(18) Watts, P.; Ponnampalam, D.; Hsu, W.; Barnes, A.; Chambers, B., The complex permittivity of multi-walled carbon nanotube–polystyrene composite films in X-band. *Chem. Phys. Lett.* **2003,** *378* (5-6), 609-614.
(19) Zhang, X.; Li, B.-W.; Dong, L.; Liu, H.; Chen, W.; Shen, Y.; Nan, C.-W., Superior Energy Storage Performances of Polymer Nanocomposites via Modification of Filler/Polymer Interfaces. *Advanced Materials Interfaces* **2018,** *5* (11), 1800096.
(20) Sharma, M.; Madras, G.; Bose, S., Size dependent structural relaxations and dielectric properties induced by surface functionalized MWNTs in poly (vinylidene fluoride)/poly (methyl methacrylate) blends. *PCCP* **2014,** *16* (6), 2693-2704.
(21) Li, Y.; Qin, F.; Estevez, D.; Wang, H.; Peng, H.-X., Interface Probing by Dielectric Frequency Dispersion in Carbon Nanocomposites. *Scientific reports* **2018,** *8* (1), 14547.
(22) Li, X.; Lu, K., Improving sustainability with simpler alloys. *Science* **2019,** *364* (6442), 733-734.
(23) Joyce, D. M.; Ouchen, F.; Grote, J. G., Re-engineering the Polymer Capacitor, Layer by Layer. *Advanced Energy Materials* **2016,** *6* (15), 1600676.
(24) Wang, B.; Liu, L.; Huang, L.; Chi, L.; Liang, G.; Yuan, L.; Gu, A., Fabrication and origin of high-k carbon nanotube/epoxy composites with low dielectric loss through layer-by-layer casting technique. *Carbon* **2015,** *85*, 28-37.
(25) Wang, J.; Shi, Z.; Mao, F.; Chen, S.; Wang, X., Bilayer Polymer Metacomposites Containing Negative Permittivity Layer for New High-k Materials. *ACS Appl Mater Interfaces* **2017,** *9* (2), 1793-1800.
(26) Zhu, J.; Shen, J.; Guo, S.; Sue, H.-J., Confined distribution of conductive particles in polyvinylidene fluoride-based multilayered dielectrics: Toward high permittivity and breakdown strength. *Carbon* **2015,** *84*, 355-364.
(27) Teirikangas, M.; Juuti, J.; Jantunen, H., Multilayer BST-COC Composite with Enhanced High Frequency Dielectric Properties. *Ferroelectrics* **2009,** *387* (1), 210-215.
(28) Li, W.; Zhang, W.; Wang, L.; Gu, J.; Chen, A.; Zhao, R.; Liang, Y.; Guo, H.; Tang, R.; Wang, C.; Jin, K.; Wang, H.; Yang, H., Vertical Interface Induced Dielectric Relaxation in Nanocomposite (BaTiO3)1-x:(Sm2O3)x Thin Films. *Scientific reports* **2015,** *5*, 11335.
(29) Yang, D.; Ruan, M.; Huang, S.; Wu, Y.; Li, S.; Wang, H.; Ao, X.; Liang, Y.; Guo, W.; Zhang, L., Dopamine and silane functionalized barium titanate with improved electromechanical properties for silicone dielectric elastomers. *RSC Advances* **2016,** *6* (93), 90172-90183.
(30) Kim, S. W.; Kim, T.; Kim, Y. S.; Choi, H. S.; Lim, H. J.; Yang, S. J.; Park, C. R., Surface modifications for the effective dispersion of carbon nanotubes in solvents and polymers. *Carbon* **2012,** *50* (1), 3-33.
(31) Datsyuk, V.; Kalyva, M.; Papagelis, K.; Parthenios, J.; Tasis, D.; Siokou, A.; Kallitsis, I.; Galiotis, C., Chemical oxidation of multiwalled carbon nanotubes. *Carbon* **2008,** *46* (6), 833-840.
(32) Spitalsky, Z.; Tasis, D.; Papagelis, K.; Galiotis, C., Carbon nanotube–polymer composites: Chemistry, processing, mechanical and electrical properties. *Prog. Polym. Sci.* **2010,** *35* (3), 357-401.
(33) Lee, H.; Dellatore, S. M.; Miller, W. M.; Messersmith, P. B., Mussel-inspired surface

chemistry for multifunctional coatings. *science* **2007,** *318* (5849), 426-430.

(34) Song, Y.; Shen, Y.; Liu, H.; Lin, Y.; Li, M.; Nan, C.-W., Enhanced dielectric and ferroelectric properties induced by dopamine-modified BaTiO3 nanofibers in flexible poly(vinylidene fluoride-trifluoroethylene) nanocomposites. *J. Mater. Chem.* **2012,** *22* (16), 8063.

(35) Mu, M.; Osswald, S.; Gogotsi, Y.; Winey, K. I., An in situ Raman spectroscopy study of stress transfer between carbon nanotubes and polymer. *Nanotechnology* **2009,** *20* (33), 335703.

(36) Dresselhaus, M. S.; Dresselhaus, G.; Saito, R.; Jorio, A., Raman spectroscopy of carbon nanotubes. *Phys. Rep.* **2005,** *409* (2), 47-99.

(37) Shah, K.; Vasileva, D.; Karadaghy, A.; Zustiak, S., Development and characterization of polyethylene glycol–carbon nanotube hydrogel composite. *Journal of Materials Chemistry B* **2015,** *3* (40), 7950-7962.

(38) Glaskova, T.; Zarrelli, M.; Aniskevich, A.; Giordano, M.; Trinkler, L.; Berzina, B., Quantitative optical analysis of filler dispersion degree in MWCNT–epoxy nanocomposite. *Compos. Sci. Technol.* **2012,** *72* (4), 477-481.

(39) Cole, K. S.; Cole, R. H., Dispersion and absorption in dielectrics I. Alternating current characteristics. *The Journal of chemical physics* **1941,** *9* (4), 341-351.

(40) Qin, F.; Brosseau, C., A review and analysis of microwave absorption in polymer composites filled with carbonaceous particles. *J. Appl. Phys.* **2012,** *111* (6), 061301.